\documentclass[prd,preprint,showpacs,superscriptaddress,amsmath]{revtex4-1}

\usepackage{longtable}
\usepackage{graphicx}
\usepackage{amsmath,amssymb}
\usepackage{color}
\usepackage{tikz}
\usepackage{units}
\usepackage{siunitx}
\usepackage{anyfontsize}
\usetikzlibrary{shapes,arrows}

\newcommand{\beq}{\begin{equation}}
\newcommand{\eeq}{\end{equation}}
\newcommand{\bdm}{\begin{displaymath}}
\newcommand{\edm}{\end{displaymath}}

\definecolor{Gray}{gray}{0.9}

\graphicspath{{./plots/}}
\begin{document}

\title{Measurement and subtraction of Schumann resonances at gravitational-wave interferometers}

\author{Michael W. Coughlin}
\affiliation{Division of Physics, Math, and Astronomy, California Institute of Technology, Pasadena, CA 91125, USA}
\author{Alessio Cirone}
\affiliation{INFN, Sezione di Genova, I-16146 Genova, Italy}
\affiliation{Universit\`a degli Studi di Genova, I-16146 Genova, Italy}
\author{Patrick Meyers}
\affiliation{School of Physics and Astronomy, University of Minnesota, Minneapolis, Minnesota 55455, USA}
\author{Sho Atsuta}
\affiliation{Department of Physics, Tokyo Institute of Technology}
\author{Valerio~Boschi}
\affiliation{European Gravitational Observatory (EGO), I-56021 Cascina, Pisa, Italy}
\author{Andrea~Chincarini}
\affiliation{INFN, Sezione di Genova, I-16146 Genova, Italy}
\author{Nelson~L.~Christensen}
\affiliation{Physics and Astronomy, Carleton College, Northfield, MN 55057, USA}
\affiliation{Artemis, Universit\'e C\^ote d'Azur, Observatoire C\^ote d'Azur, CNRS, CS 34229, F-06304 Nice Cedex 4, France}
\author{Rosario~De~Rosa}
\affiliation{INFN, Sezione di Napoli, Complesso Universitario di Monte S.Angelo, I-80126 Napoli, Italy}
\affiliation{Universit\`a di Napoli 'Federico II', Complesso Universitario di Monte S.Angelo, I-80126 Napoli, Italy}
\author{Anamaria Effler}
\affiliation{LIGO Livingston Observatory, Livingston, LA 70754 USA}
\author{Irene Fiori}
\affiliation{European Gravitational Observatory (EGO), I-56021 Cascina, Pisa, Italy}
\author{Mark Go{\l}kowski}
\affiliation{Department of Electrical Engineering, University of Colorado Denver, Denver, CO 80204, USA}
\author{Melissa Guidry} 
\affiliation{Department of Physics, College of William and Mary, Williamsburg, VA 23185, USA}
\author{Jan Harms}
\affiliation{Gran Sasso Science Institute (GSSI), I-67100 L’Aquila, Italy}
\affiliation{INFN, Laboratori Nazionali del Gran Sasso, I-67100 Assergi, Italy}
\author{Kazuhiro Hayama}
\affiliation{KAGRA Observatory, Institute for Cosmic Ray Research, University of Tokyo, Japan}
\author{Yuu Kataoka}
\affiliation{Department of Physics, Tokyo Institute of Technology}
\author{Jerzy Kubisz}
\affiliation{Astronomical Observatory, Jagiellonian University, Krakow, Poland}
\author{Andrzej Kulak}
\affiliation{AGH University of Science and Technology, Department of Electronics, Krakow, Poland}
\author{Michael Laxen}
\affiliation{LIGO Livingston Observatory, Livingston, LA 70754 USA}
\author{Andrew Matas}
\affiliation{School of Physics and Astronomy, University of Minnesota, Minneapolis, Minnesota 55455, USA}
\author{Janusz Mlynarczyk}
\affiliation{AGH University of Science and Technology, Department of Electronics, Krakow, Poland}
\author{Tsutomu Ogawa}
\affiliation{Earthquake Research Institute, University of Tokyo, Japan}
\author{Federico Paoletti}
\affiliation{INFN, Sezione di Pisa, I-56127 Pisa, Italy}
\affiliation{European Gravitational Observatory (EGO), I-56021 Cascina, Pisa, Italy}
\author{Jacobo Salvador}
\affiliation{Observatorio Atmosferico de la Patagonia Austral, OAPA UNIDEF (MINDEF-CONICET), Río Gallegos, Argentina}
\author{Robert Schofield}
\affiliation{University of Oregon, Eugene, OR 97403, USA}
\author{Kentaro Somiya}
\affiliation{Department of Physics, Tokyo Institute of Technology}
\author{Eric~Thrane}
\affiliation{School of Physics and Astronomy, Monash University, Clayton, Victoria 3800, Australia}

\begin{abstract}
Correlated magnetic noise from Schumann resonances threatens to contaminate the observation of a stochastic gravitational-wave background in interferometric detectors. In previous work, we reported on the first effort to eliminate global correlated noise from the Schumann resonances using Wiener filtering,  demonstrating as much as a factor of two reduction in the coherence between magnetometers on different continents. 
In this work, we present results from dedicated magnetometer measurements at the Virgo and KAGRA sites, which are the first results for subtraction using data from gravitational-wave detector sites.
We compare these measurements to a growing network of permanent magnetometer stations, including at the LIGO sites.
We show how dedicated measurements can reduce coherence to a level consistent with uncorrelated noise.
We also show the effect of mutual magnetometer attraction, arguing that magnetometers should be placed at least one meter from one another.
\end{abstract}

\maketitle

\section{Introduction}

A detection of a stochastic gravitational-wave background (SGWB) would be a significant result for gravitational-wave astronomy, having far-reaching implications for cosmology and astrophysics.
One potential method for detecting a SGWB is to use a network of ground-based, second-generation interferometric gravitational-wave detectors, which currently consists of Advanced LIGO \cite{aligo} and Advanced Virgo \cite{avirgo}. 
A SGWB from compact binary coalescences is potentially detectable by the time second-generation detectors reach design sensitivity~\cite{AbEA2017h}. Backgrounds from pulsars, magnetars, core-collapse supernovae, and various physical processes in the early universe are all possible as well \cite{AbEA2009,AbEA2012s,AbEA2016b}, but their expected amplitudes are not as well constrained as the expected background due to compact binary coalescences.
The potential for the contamination of searches for a SGWB is strong due to potential correlated environmental noise between detectors \cite{TCS2013,TCS2014, LSC2007a,LSC2007b,AbEA2012s,AaEA2014}, which would result in a systematic error in the searches. 
A related concern exists in searches for transient sources of gravitational waves, such as due to correlated magnetic transients from storms~\cite{KoBi2017}.

\begin{figure}[t]
\centering
\hspace*{-0.5cm}
\includegraphics[width=3.5in]{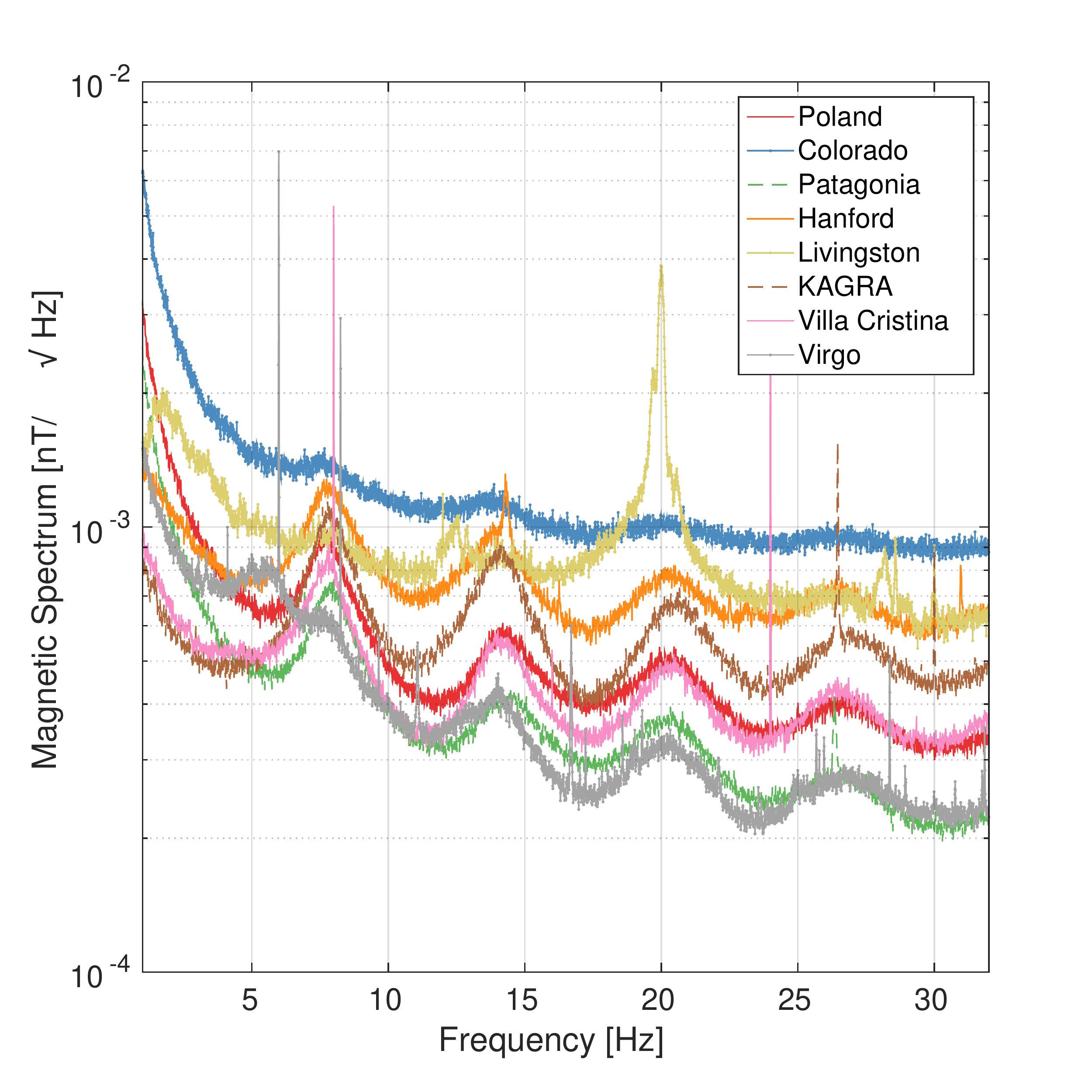}
\caption{Median power spectral density of the North-South Poland, North-South Colorado, North-South Villa Cristina, North-South Patagonia, KAGRA X-arm direction, LIGO Hanford X-arm direction, and LIGO Livingston X-arm direction magnetometers. These are computed using 128\,s segments. In addition to the sharp instrumental line features in the Villa Cristina magnetometer at 8\,Hz and 24\,Hz, the Schumann resonances are visible in all of the magnetometers. The 20~Hz line at LIGO Livingston is likely due to power lines which cross the site on the Y-arm.}
\label{fig:psd}
\end{figure}

One such source of correlated noise is global electromagnetic fields such as the Schumann resonances~\cite{Schumann1}, which induce forces on magnets or magnetically susceptible materials in the test-mass suspension system.
Schumann resonances arise in the Earth-ionosphere waveguide from the tiny attenuation of extremely low frequency electromagnetic waves.
The broadly peaked Schumann resonances at 8, 14, 21, 27, and 32\,Hz  are potentially problematic for the SGWB searches. The power spectral density (PSD) showing these features from magnetometer data for the sites of interest in this analysis are shown in Figure~\ref{fig:psd}.
While the primary peak is below the seismic wall of the gravitational-wave detectors at 10\,Hz and therefore will not affect sensitivity for current detectors, the secondary and tertiary harmonics at 14\,Hz and 20\,Hz respectively could be limiting noise sources.

In previous work, we carried out a demonstration of Wiener filtering with a goal of reducing the coherence between widely separated magnetometers (serving as proxies for gravitational-wave detectors) \cite{CoCh2016}.
We used previously deployed magnetometers, which allowed us access to instruments with superb sensitivity, located in very magnetically quiet locations. 
In addition, at Virgo, a temporary station was created at Villa Cristina, which is a magnetically quiet site 12.72\,km southwest from Virgo. 

In that paper, we argued that it would be beneficial to measure the coherence between quiet magnetometers stationed near gravitational-wave detectors.
In this present analysis, we use these magnetometers in addition to dedicated measurements at the gravitational-wave detectors to perform subtraction with realistic levels of local magnetic noise of correlated magnetic noise in gravitational-wave detectors.
In the following, we use these magnetometers as a proxy for a gravitational-wave interferometer strain channel. 
We show that magnetic correlations are significant over the entire Schumann band.
We also demonstrate that using co-located and co-aligned magnetometers results in cleaned data consistent with noise.

\section{Magnetometer stations}

\begin{figure*}[t]
\hspace*{-0.5cm}
 \includegraphics[width=6in]{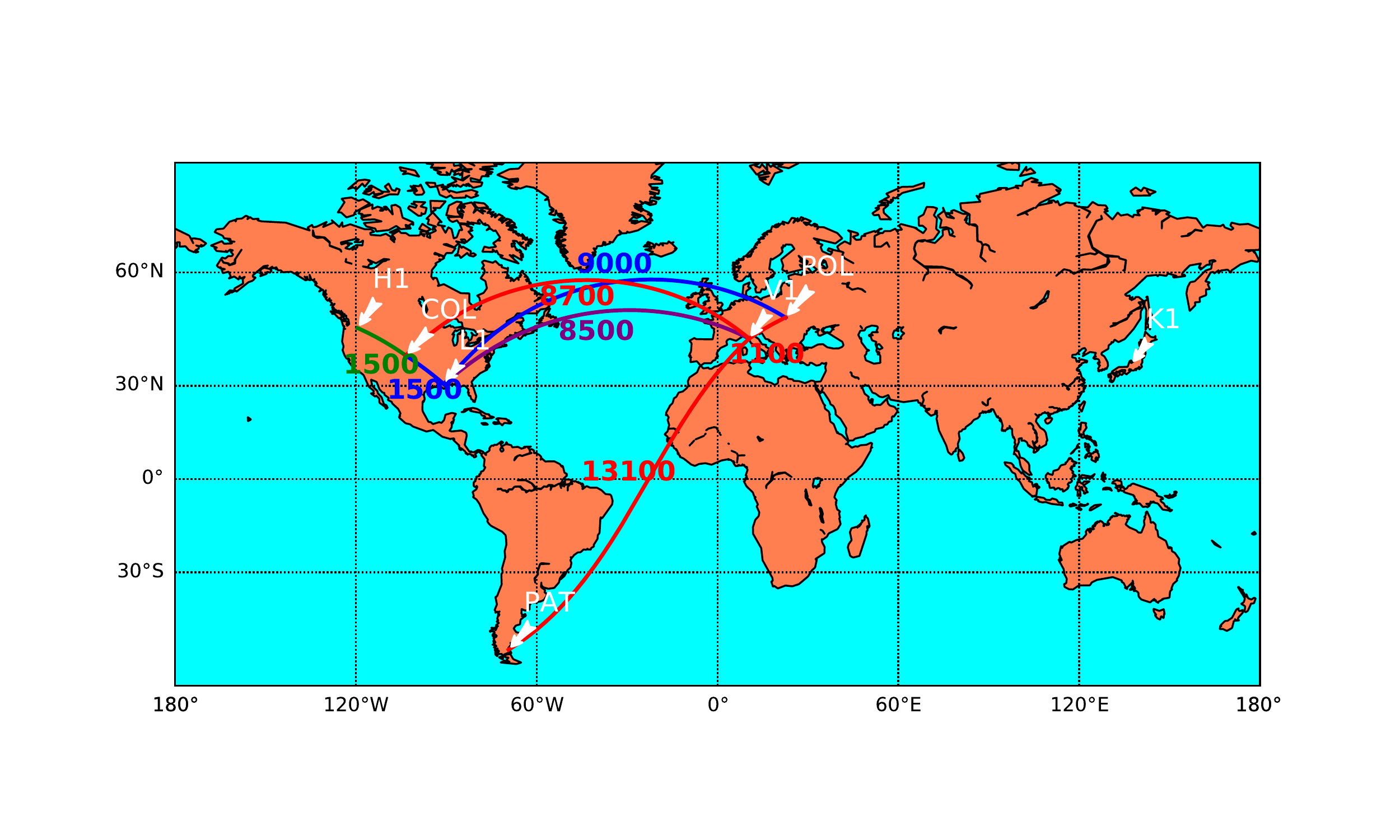}
 \caption{The plot shows the location of the 7 magnetometer stations with available data during this study (Here V1 and Villa Cristina are treated as being co-located). We show the networks used in the Wiener filtering analysis, represented as green, blue and red lines with the distance in kilometers shown. The purple line is a pair (Villa Cristina and LIGO Livingston) which is used in both networks.}
 \label{fig:map}
\end{figure*}

\begin{figure*}[t]
\hspace*{-0.5cm}
 \includegraphics[width=5in]{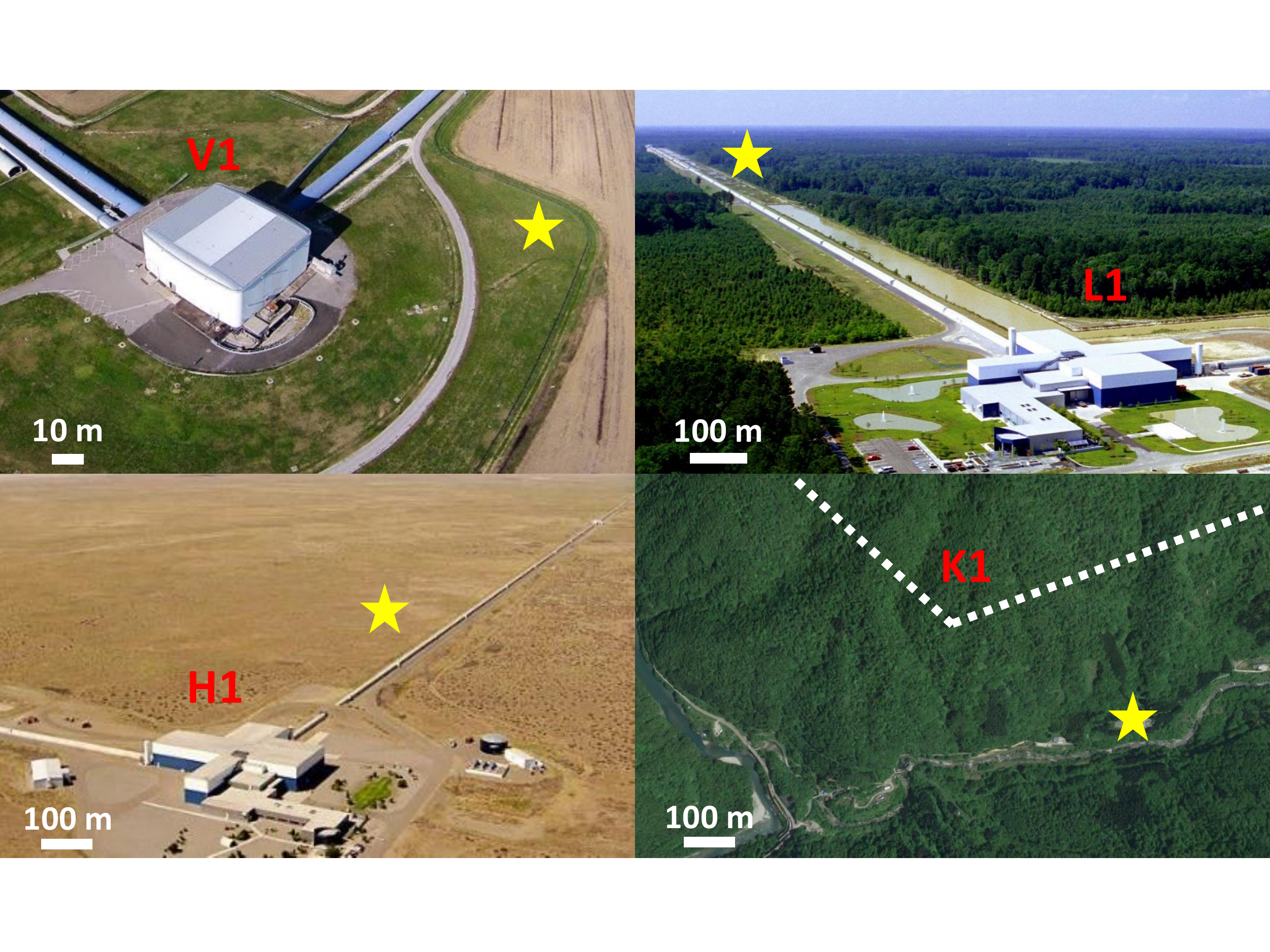}
 \caption{Locations of the magnetometers at Virgo, Livingston, Hanford and KAGRA. Their coordinates in the interferometer system (x,y) 
are: $V1 = (80,-72)\,\textrm{m};\ L1 = (120,3000)\,\textrm{m};\ H1 = (1030,195)\,\textrm{m};\ K1 = (400,-600)\,\textrm{m}$.}
 \label{fig:pictures}
\end{figure*}

\begin{table}[]
{\fontsize{0.3cm}{0.3cm}\selectfont
\begin{tabular}{ | p{3.3cm} | p{2.3cm} | p{2.7cm} | p{1.5cm} | p{3.8cm} |} \hline
   & Location & Orientation & Type & Suppl. info \\ \hline
   Hanford (H1) & \ang{46}27'42.2"N \ang{119}25'03.6"W & X, Y arms & LEMI-120 & permanent \\ \hline    
   Livingston (L1) & \ang{30}32'12.9"N \ang{90}45'57.5"W & X, Y arms (LIGO1, LIGO2) & LEMI-120 & permanent \\ \hline    
   Virgo (V1) & \ang{43}37'54.7"N \ang{10}°30'20.1"E & NS, EW & MFS-06 & permanent (8-26 Aug 2017) \\ \hline    
   Villa Cristina (VC) & \ang{43}32'22.2"N, \ang{10}24'36"E & NS, EW & MFS-06 & temporary (20-22 July and 22-24 Nov 2016) \\ \hline
   KAGRA 1 (K1) & \ang{36}24'33.5"N \ang{137}18'39.4"E & NS, EW, Vertical & MFS-06 & temporary (20-22 July 2016) \\ \hline
   KAGRA 2 & \ang{36}24'42"N, \ang{137}18'18"E & NS, EW & MFS-06 & temporary \\ \hline
   Hylaty station (POL) & \ang{49.2}N, \ang{22.5}E & NS, EW & AAS1130 & permanent \\ \hline
   Hugo station (COL) & \ang{38.9}N, \ang{103.4}W & NS, EW & AAS1130 & permanent \\ \hline
   Patagonia station (PAT) & \ang{51.5}S, \ang{69.3}W & NS, EW & AAS1130 & permanent \\ \hline
\end{tabular}
\caption{Properties of all the magnetic antennas occurring in the text. We note here that LIGO1 and LIGO2 are noted for pointers in later figures; there are the same number of LEMI magnetometers at Hanford and Livingston.}
\label{magtable}}
\end{table}

In this study, we use a variety of permanent and temporary extremely low frequency (ELF) magnetometer installations (Table~\ref{magtable}). These magnetometers have sensitive bands of 3-300\,Hz, with a sensitivity of $\approx 0.015 \textrm{pT}/\textrm{Hz}^{1/2}$ at 14\,Hz \cite{GaNi2015}.
Three of the permanent installations are part of the WERA project \footnote{http://www.oa.uj.edu.pl/elf/index/projects3.htm}, the Hylaty station in the Bieszczady Mountains in Poland, the Hugo Station located in the Hugo Wildlife Area in Colorado, USA, and the Patagonia station located in Rio Gallegos in Patagonia, Argentina \cite{KuKu2014}.
These stations contain two magnetometers oriented North-South and East-West.

Three more permanent stations, one each at the LIGO Hanford, LIGO Livingston, and Virgo interferometers were established.
The LIGO magnetometers are LEMI-120's \footnote{http://www.lemisensors.com/?p=245} while the Virgo magnetometers are MFS-06 by Metronix \footnote{https://www.geo-metronix.de/mtxgeo/index.php/mfs-06e-overview}, which are broadband induction coil magnetometers designed to measure variations of the Earth's magnetic field.
These witness sensors are placed far enough from the gravitational-wave detectors so as to not be sensitive to local magnetic noise but close enough to measure approximately the same Schumann resonances as the detectors do.
The LIGO sensors are aligned with the X- and Y-arms of the detectors, while the Virgo sensors are aligned North-South and East-West.

In addition, temporary measurements are also made. 
Dedicated measurements both inside and outside the mine of KAGRA site at Kamioka and at Villa Cristina near Virgo were performed to supplement the permanent stations. 
Data at Villa Cristina was taken with both North-South and East-West facing magnetometers (MFS-06 by Metronix).
The measurements were made between July 20-22, 2016, and additional measurements between November 22-24, 2016.
At KAGRA, two temporary stations, both outside the mine about 30\,m apart, were created between July 20-22, 2016 with North-South, East-West, and vertical magnetometers (MFS-06 by Metronix).

Figure~\ref{fig:map} shows the location of magnetometer stations that had data during this study. The colors indicate the networks used in the following analysis, as well as the distances in kilometers between the pairs.
For example, for the Virgo network, the distances between this sensor and the witness sensors are approximately 1100\,km for Poland Hylaty, and 8700\,km for Colorado Hugo, and 13,100\,km for Patagonia.
Figure~\ref{fig:pictures} shows the location on the individual sites for the magnetometers relative to the interferometer vertex for the LIGO and Virgo detectors, and relative to the planned location for KAGRA.

\section{Coherence Measurements}

\begin{figure*}[t]
\centering
\hspace*{-0.5cm}
\includegraphics[width=2.0in]{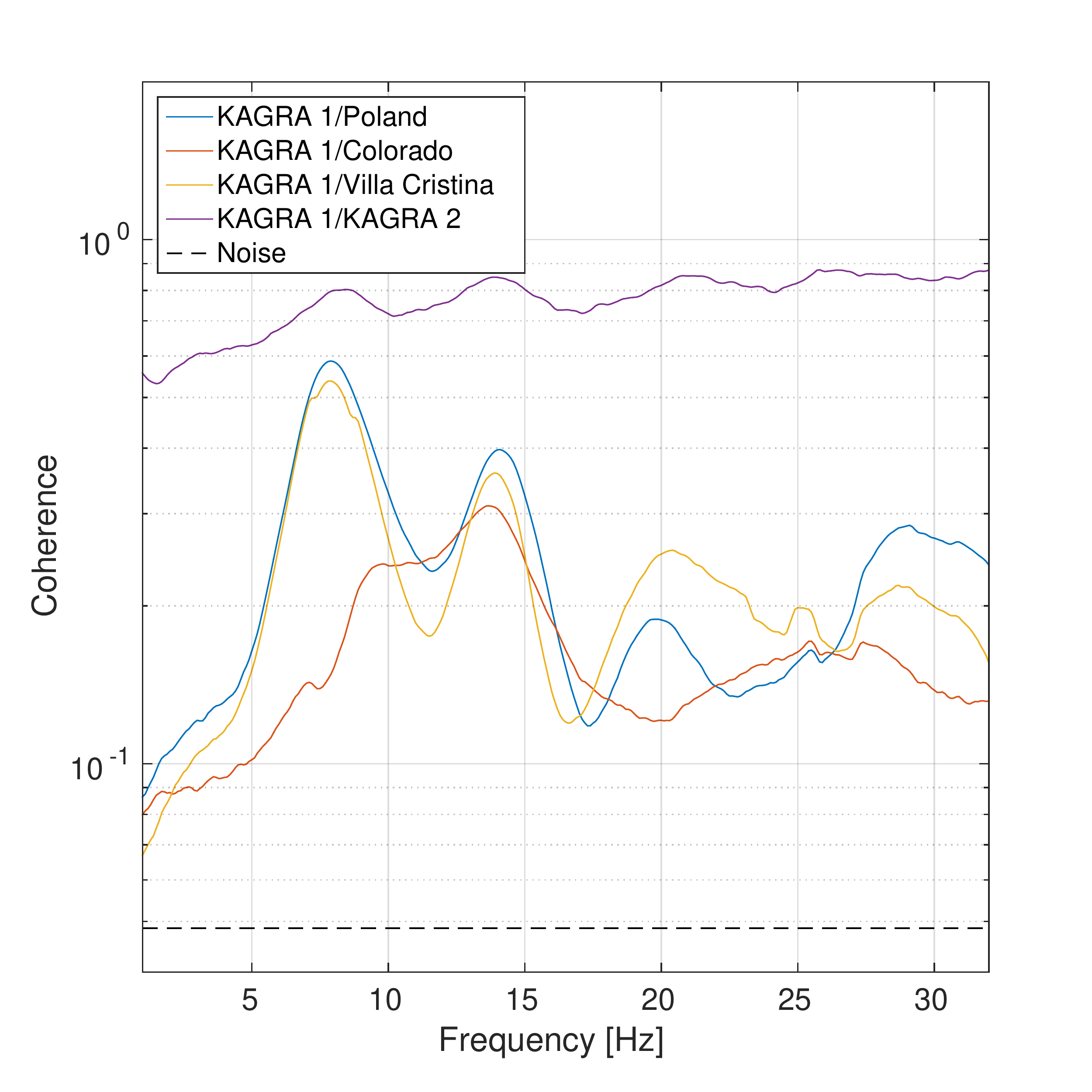}
\includegraphics[width=2.0in]{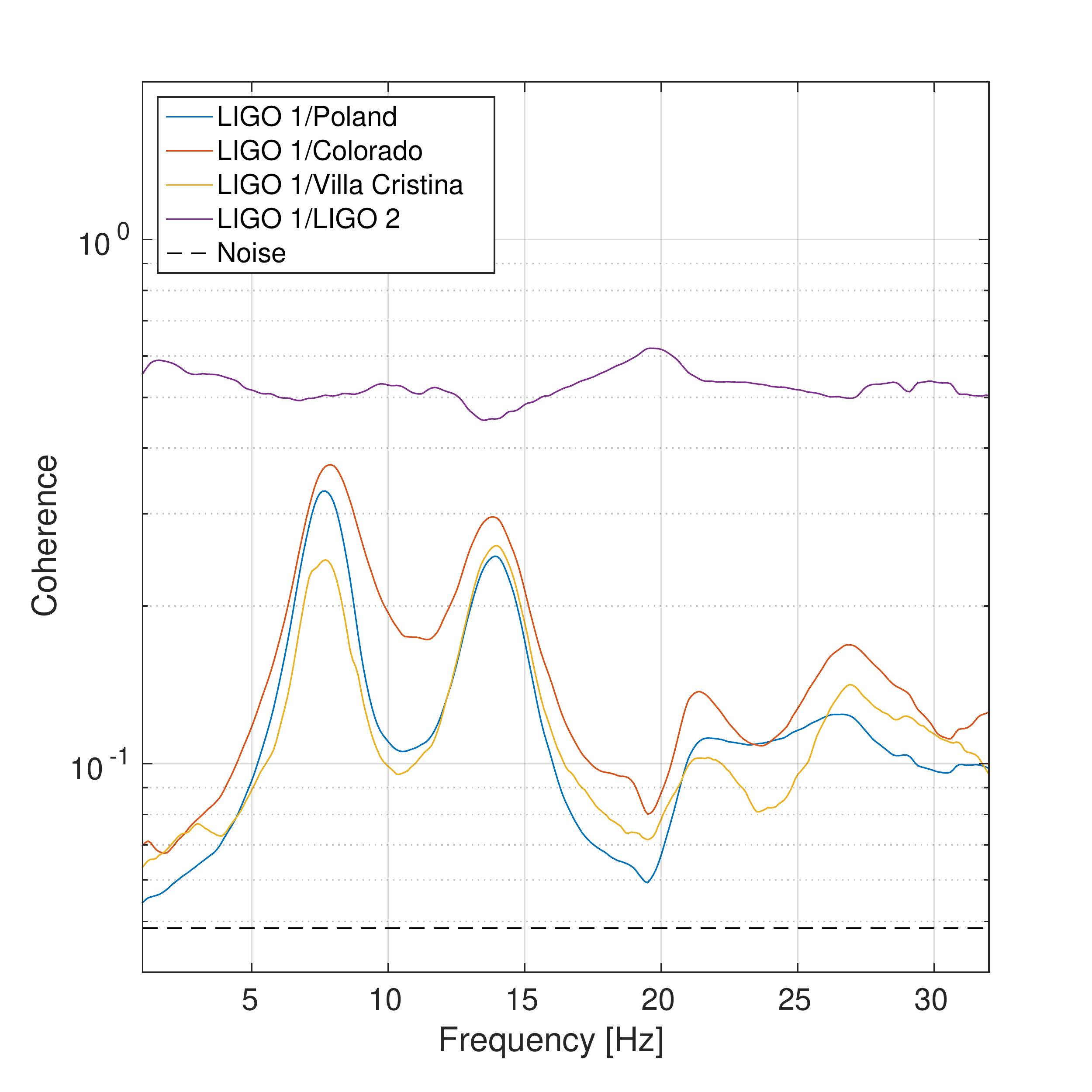}
\includegraphics[width=2.0in]{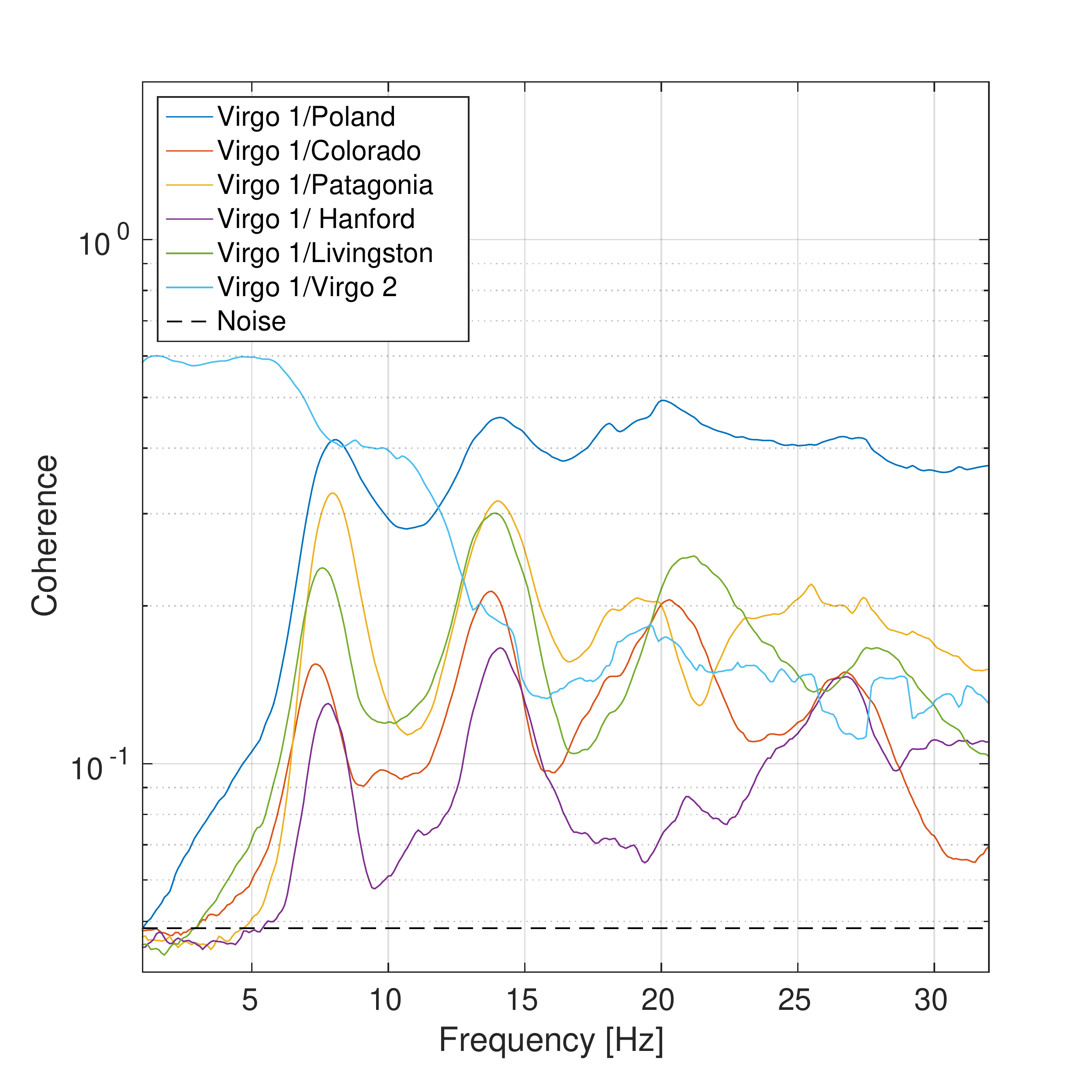}
KAGRA \hspace{1.2in} LIGO Livingston \hspace{1.2in} Virgo
\caption{On the left is the coherence between KAGRA and the North-South Poland, North-South Colorado, and Villa Cristina magnetometers over 2 days of coincident data. In this analysis, KAGRA 1 is the magnetometer located outside of the cave, and KAGRA 2 is the in-cave magnetometer. In addition, we plot the expected correlation given Gaussian noise. In the middle is the same between LIGO Livingston X-arm direction and those stations. In this analysis, LIGO 1 is the X-arm direction magnetometer, and LIGO 2 is the Y-arm direction magnetometer. The correlation between the two LIGO magnetometers is dominated by local noise, which means a Wiener filter using one of these to clean the other may not reduce correlations due to Schumann resonances, but is still useful to include in the filtering because it can increase sensitivity of the target channel by removing local disturbances. On the right is the coherence between Virgo North-South magnetometer and the North-South Poland, North-South Colorado, North-South Patagonia, LIGO Livingston X-arm direction, and LIGO Hanford X-arm direction magnetometers over a week of coincident data. In this analysis, Virgo 1 is the North-South magnetometer, and Virgo 2 is the East-West magnetometer}
\label{fig:coh}
\end{figure*}

We begin by characterizing the level of correlation between magnetometers, which will be important for the efficacy of possible subtraction using Wiener filtering.
One metric for measuring the correlation between magnetometers is the coherence $c(f)$
\begin{equation}
c(f) = \frac{|\overline{\tilde{s}_1(f) \tilde{s}_2^*(f)}|}{\overline{|\tilde{s}_1(f)|} \overline{|\tilde{s}_2(f)}|},
\label{eq:coh}
\end{equation}
where $\tilde{s}_1(f)$ and $\tilde{s}_2(f)$ are the Fourier transforms of the two channels, and $^*$ indicates complex conjugation. 

Figure~\ref{fig:coh} shows the coherence of the North-South Poland, North-South Colorado, Villa Cristina, and one of the KAGRA magnetometers with another of the KAGRA magnetometers. Clear peaks are visible in all pairs. 
The two spatially co-located and co-aligned KAGRA magnetometers provide an excellent test-bed for the best subtraction possible.
They show broadband coherence between 0.5-0.9, which will result in excellent subtraction.
The other magnetometer correlations result in coherence that peaks at about 0.6 for the dominant harmonic.
This is similar to what was achieved from the last data set \cite{CoCh2016}. 
The middle of Figure~\ref{fig:coh} shows correlation between the LIGO Livingston X-arm direction magnetometer with the same set of magnetometers and the LIGO Livingston Y-arm direction magnetometer. We again see the curve with the highest coherence is the one dominated by the colocated magnetometers, and peaks corresponding to Schumann resonances in the other curves. While the colocated LIGO curve is clearly dominated by local magnetic noise, as opposed to Schumann resonances, it is important to include this channel in the subtraction network as it will help remove local noise in the target channel and aid in the ability to subtract the global magnetic noise from the target channel as well.
Finally, the right of Figure~\ref{fig:coh} shows the same analysis for a Virgo magnetometer. The results are qualitatively similar to the LIGO case.

\section{Search for a low-noise location at Virgo} 

\begin{figure}[t]
\centering
\hspace*{-0.5cm}
 \includegraphics[width=3.0in]{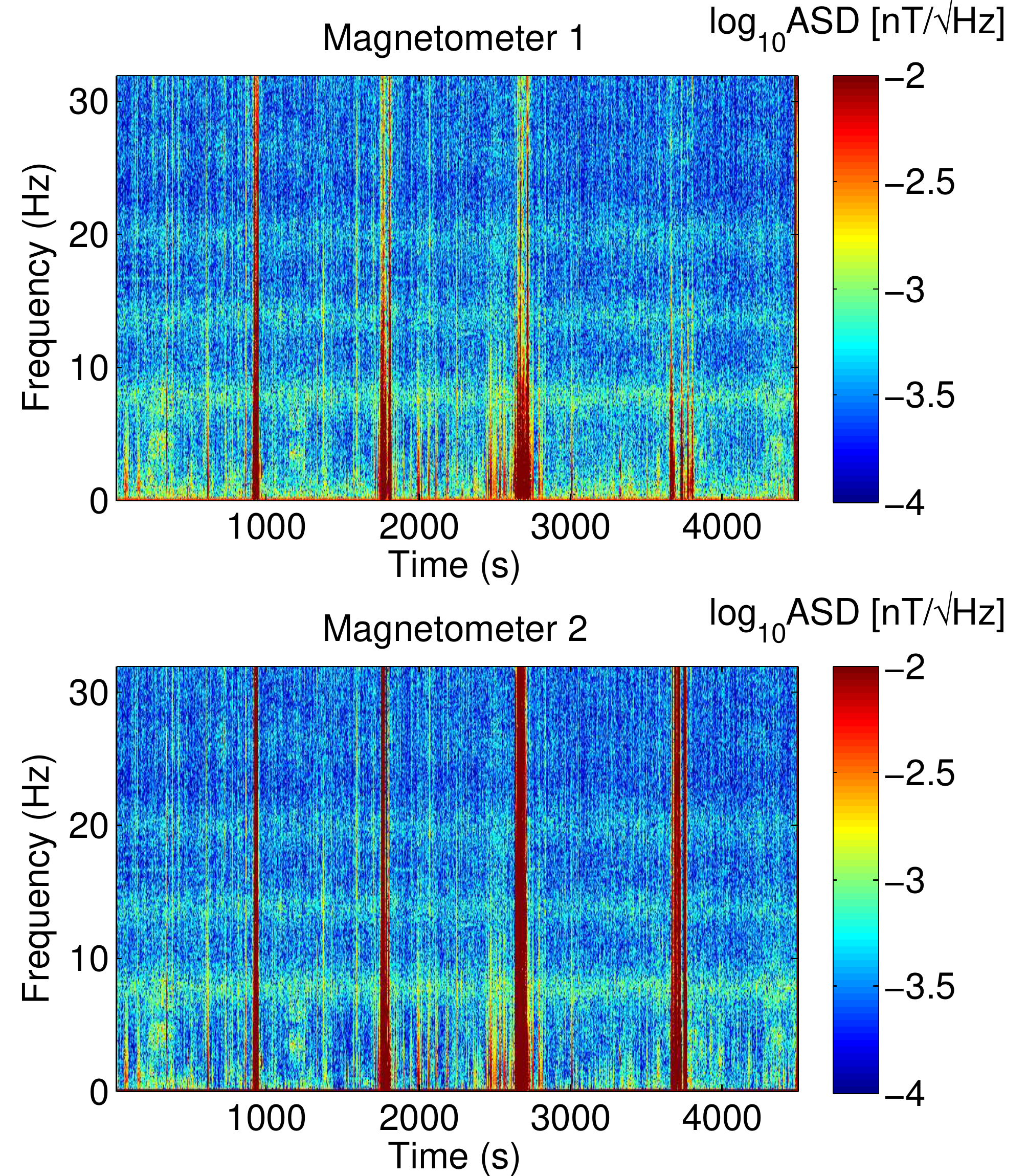}
 \includegraphics[width=3.5in]{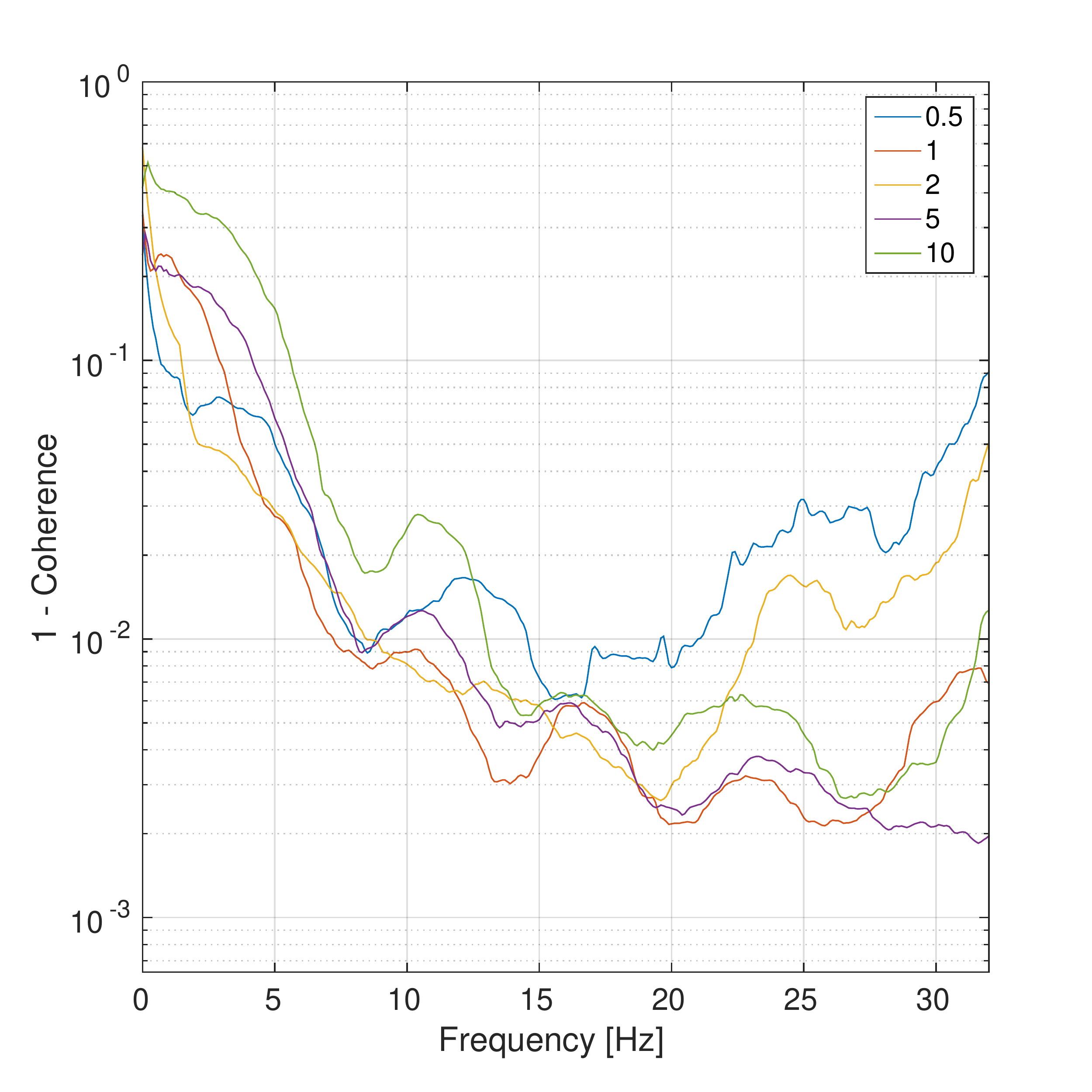}
 0.5\,m \hspace{0.1in} 1\,m \hspace{0.1in} 2\,m \hspace{0.1in} 5\,m \hspace{0.1in} 10\,m \hspace{4in}
 \caption{Spectrograms (left) and Coherence (right) of co-located and co-aligned magnetometers at Virgo at distances of 0.5\,m, 1\,m, 2\,m, 5\,m, and 10\,m.}
 \label{fig:correlation}
\end{figure}

A very preliminary measurement involved two co-located and co-aligned magnetometers at Villa Cristina, deep in the countryside near Virgo. 
The test was carried out at varying distances between the magnetometers (0.5\,m, 1\,m, 2\,m, 5\,m, and 10\,m) and it represents the ideal situation, as it seeks to remove both Schumann and local
noise simultaneously, leaving only some possible contributions coming from a misalignment or magnetometer non-linearities. In this way, it acts as a sanity check for potential noise subtraction.
In addition, it should be possible to detect any potential magnetometer mutual couplings.
Because the magnetometers use an induction coil, it is possible that two nearby magnetometers (or a magnetometer very near to a 
gravitational-wave detector) influence one another through magnetic coupling.
Therefore, it is difficult to disentangle potential coherence between nearby magnetometers due to either a mutual feedback effect or a true measurement of the far magnetic
field. Moreover, as distance increases, the magnetometers could be sensitive to near field sources, hence the necessity to be in a magnetically quiet environment. 
The results of this test are presented in Figure~\ref{fig:correlation}. The time-frequency plots on the left include five different measurements, each at increasing magnetometer 
mutual distance. The noise features between each measurement are simply due to the act of moving one of the two probes away. The Schumann peaks are visible in all measurements
in both magnetometers. On the right of Figure~\ref{fig:correlation}, we plot the coherence of the magnetometers.
We show that coherence is maximized for the probes when they are 1\,m apart. 

\begin{figure*}[t]
\centering
\hspace*{-0.5cm}
\includegraphics[width=2.9in]{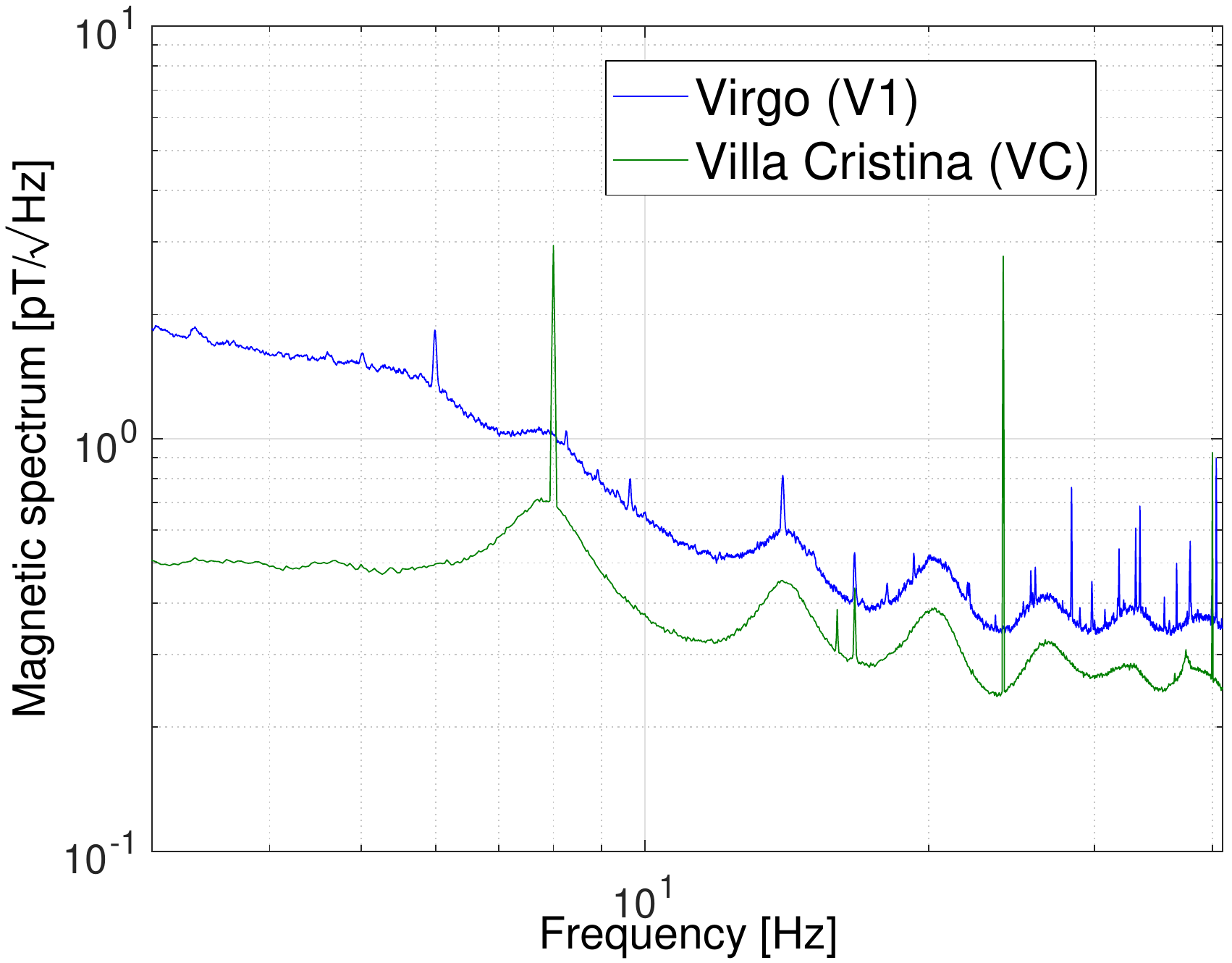}
\includegraphics[width=2.9in]{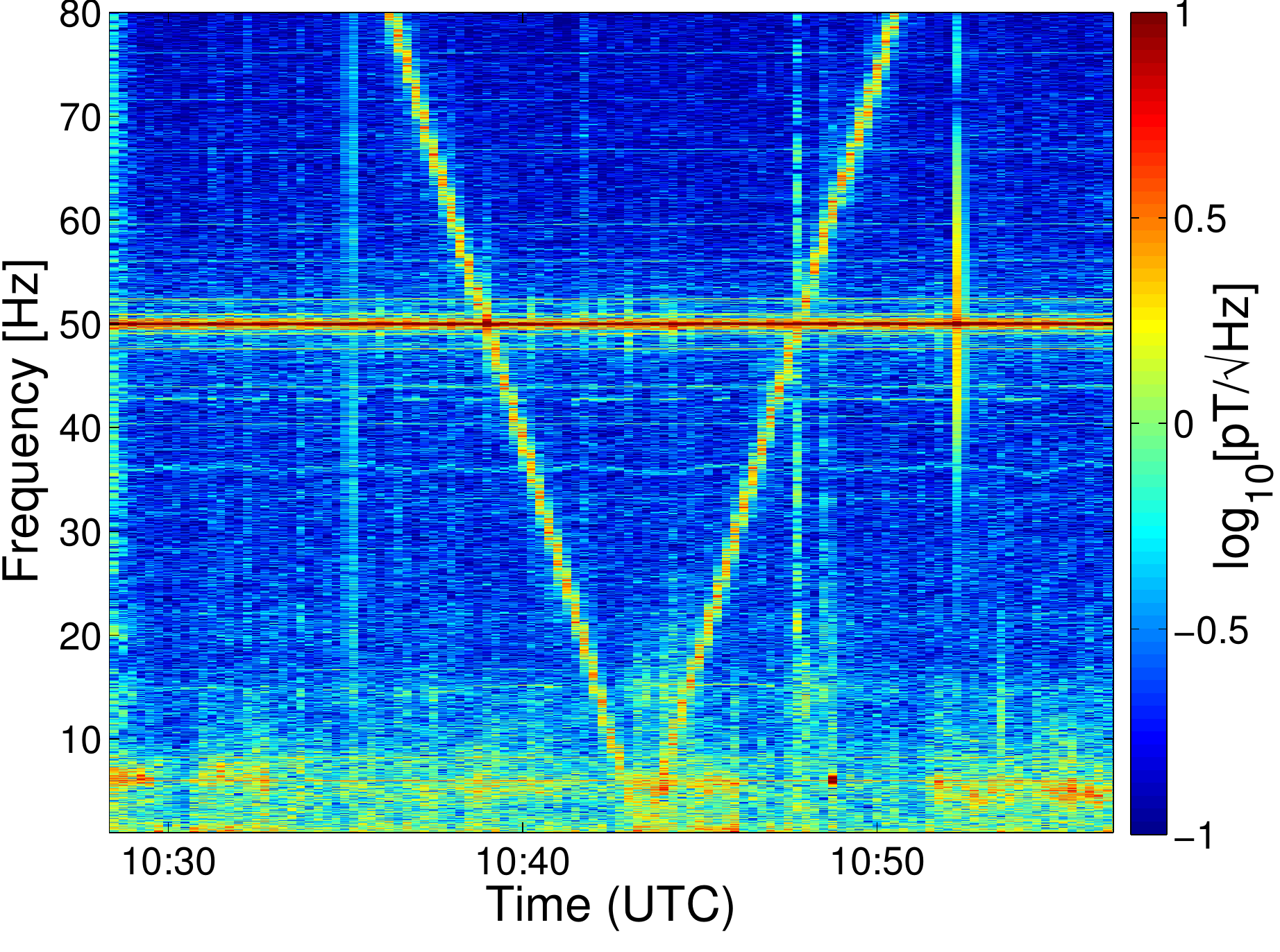}
\caption{On the left is a comparison of spectra from the  Virgo low-noise location and the one in Villa Cristina (located about $13$\,km from Virgo). 
They are both acquired with a magnetometer MFS-06 by Metronix oriented North-South.
On the right is an example of the V-shaped data transient in Virgo magnetometers data.}
\label{fig:Virgo-VC}
\end{figure*}

In addition, with the aim of building a permanent measurement station of the Schumann resonances within Virgo boundaries, we also performed an extended magnetic field survey 
looking for quiet locations along the interferometer arms and around the main buildings. One of the quietest locations was North-East of the Central Building. 
At that location, we buried two orthogonal magnetometers at $1.5$~m mutual distance for a measurement in coincidence with the first joint Advanced LIGO and Advanced Virgo observing run in 
August 2017. The sensors have been connected to a Centaur Digital Recorder by Nanometrics \footnote{https://scientists.virgo-gw.eu/EnvMon/List/Trillium/Centaur}
that acquires and stores the data at 200 Hz sampling frequency. The acquisition is synchronized with GPS time using an external antenna. 
Schumann resonances are seen clearly in the data, as shown in the blue trace on the left of Figure~\ref{fig:Virgo-VC}, where the first six Schumann peaks can be seen. 
The figure also shows a spectrum taken at Villa Cristina from the temporary measurement described previously (green line), which emphasizes the substantial difference between the two sites.
The difference is maximized below $10$~Hz, mainly due to the distinctive natural/anthropogenic seismic activity
of the locations.
In addition, there are spectral features still to be better understood, which could be related to the data acquisition system or to nearby power lines.

We also give an example of the type of undesirable transient features that can be found in the Virgo magnetometers.
Some of the most common ones are caused by nearby storms
or seismic activity in the surrounding area; others may result from probe failures or bad tuning. 
An example of the latter is a characteristic V-shape in time-frequency plots, and can be found in Figure~\ref{fig:Virgo-VC}.
We suspect that the ``chopper mode'' feature of the Virgo magnetometers, which uses a free-running local oscillator whose frequency could be temperature dependent, is the cause, because the features are evident 2-4 times per day during daylight hours.
In addition, its intensity changes between instances.
It is likely this can be mitigated by switching off the chopper mode, although this solution will likely result in some loss of magnetometer sensitivity.

\section{Wiener Filtering}

\begin{figure}[t]
\centering
\hspace*{-0.5cm}
 \includegraphics[width=3.5in]{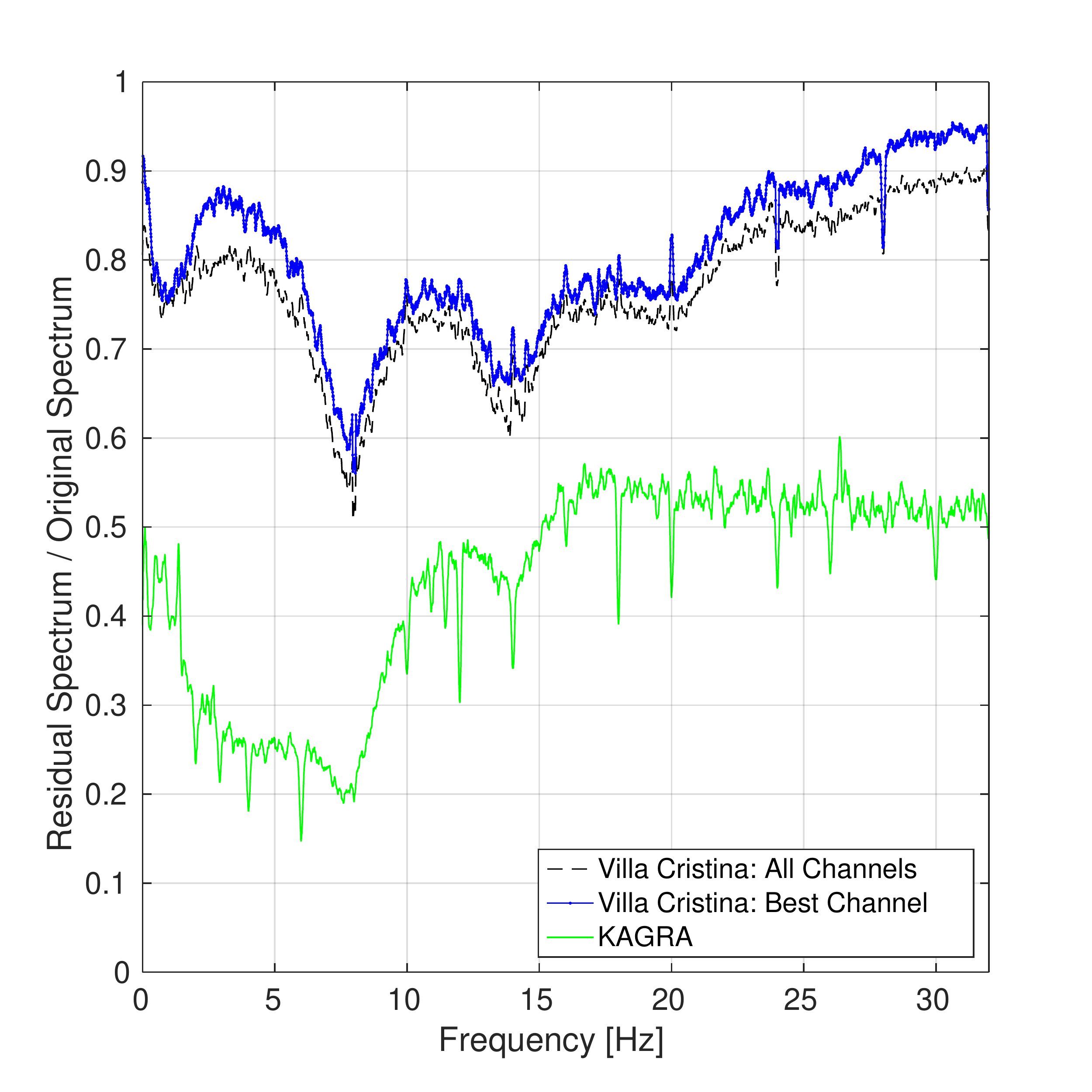}
 \caption{Ratio of the auto power spectral density before and after Wiener filter subtraction. In the first example, we use one KAGRA magnetometer as the target sensor and use the second KAGRA magnetometer as the witness sensor. In the second example, we compute the ratio of the auto power spectral density before and after Wiener filter subtraction using the Villa Cristina magnetometer as the target sensor.
We show subtraction using only the sensor with the highest coherence in each frequency bin as well as using all available sensors as witnesses, which shows improvement.
 }
 \label{fig:subtraction_auto}
\end{figure}

\begin{figure}[t]
\centering
\hspace*{-0.5cm}
 \includegraphics[width=2.0in]{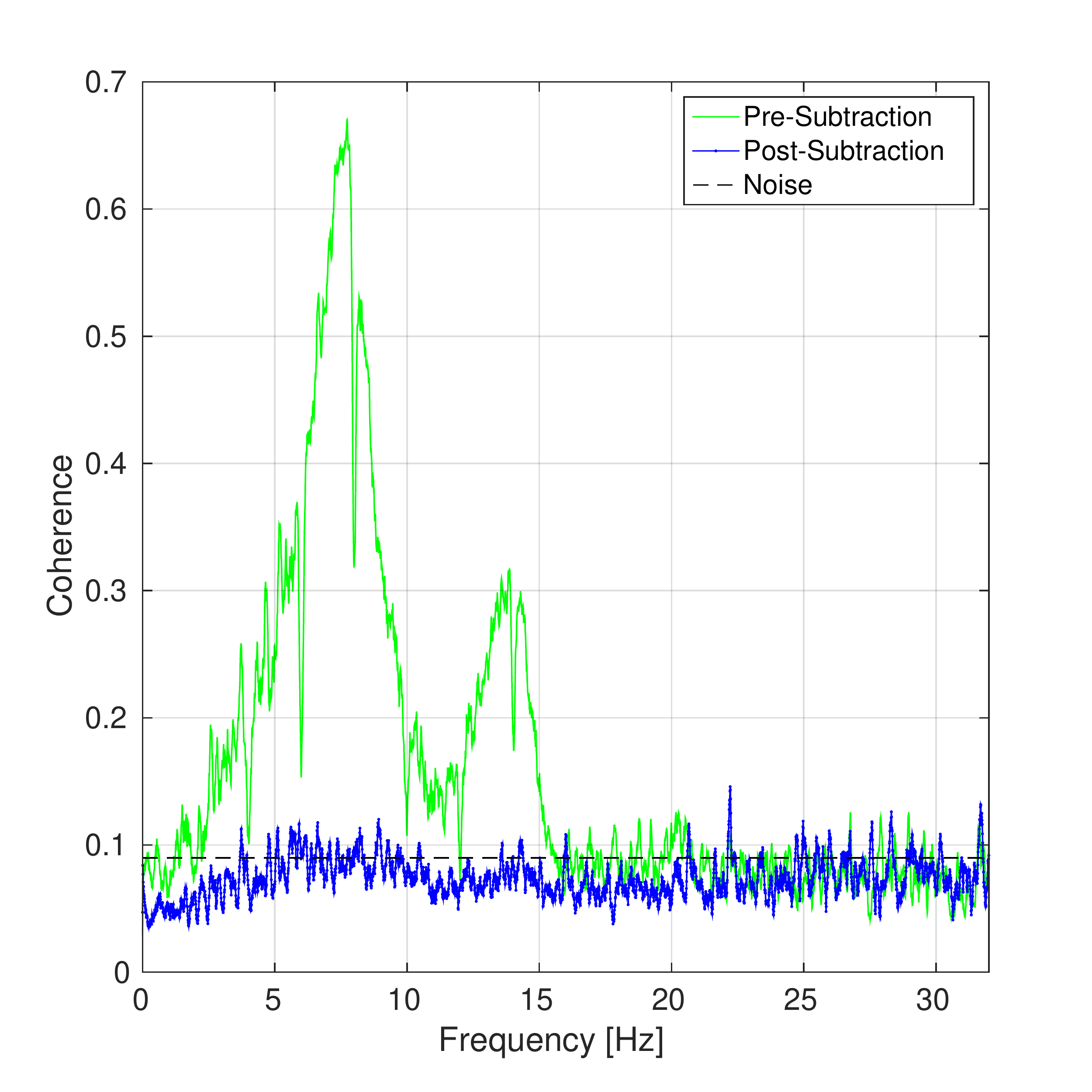}
 \includegraphics[width=2.0in]{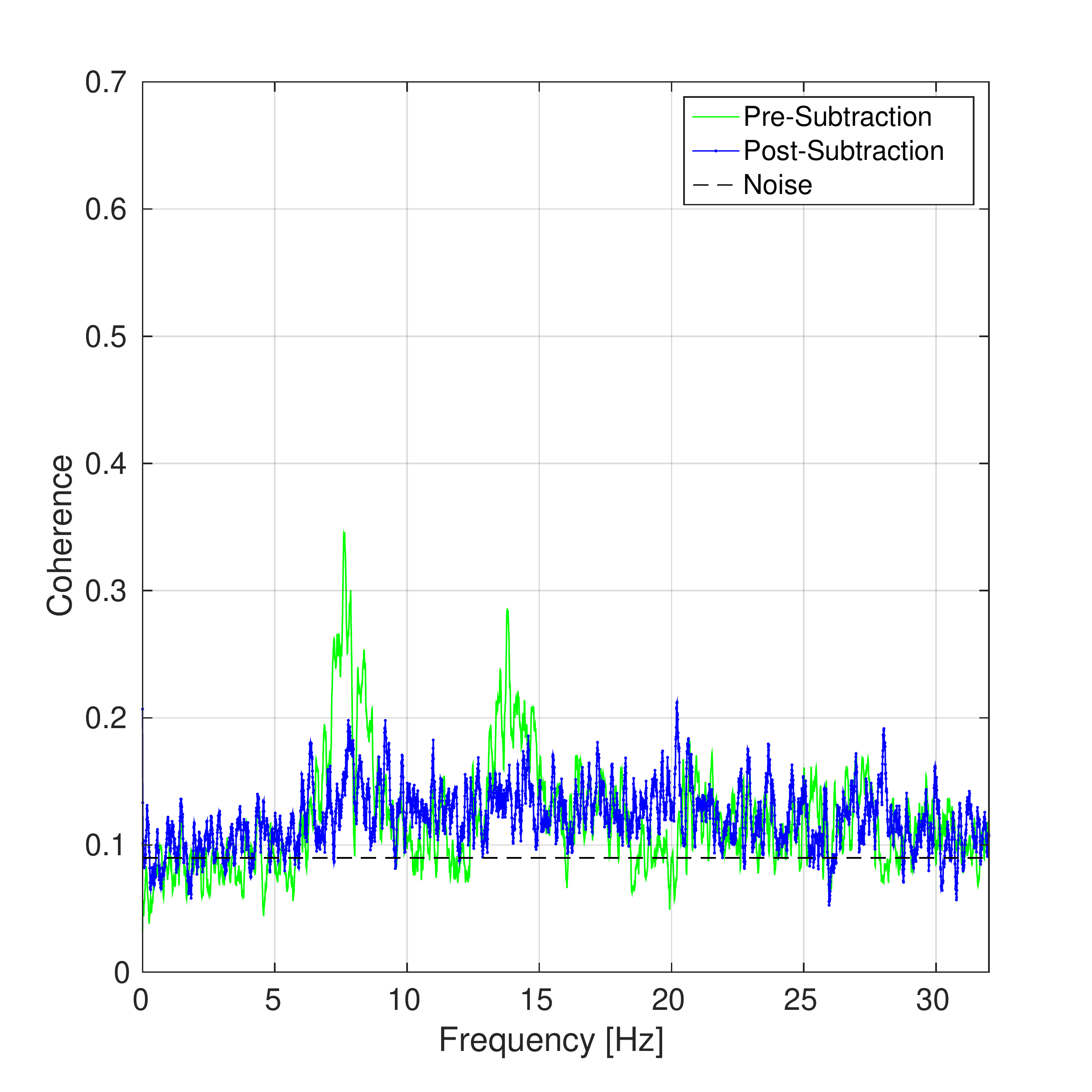}
 \includegraphics[width=2.0in]{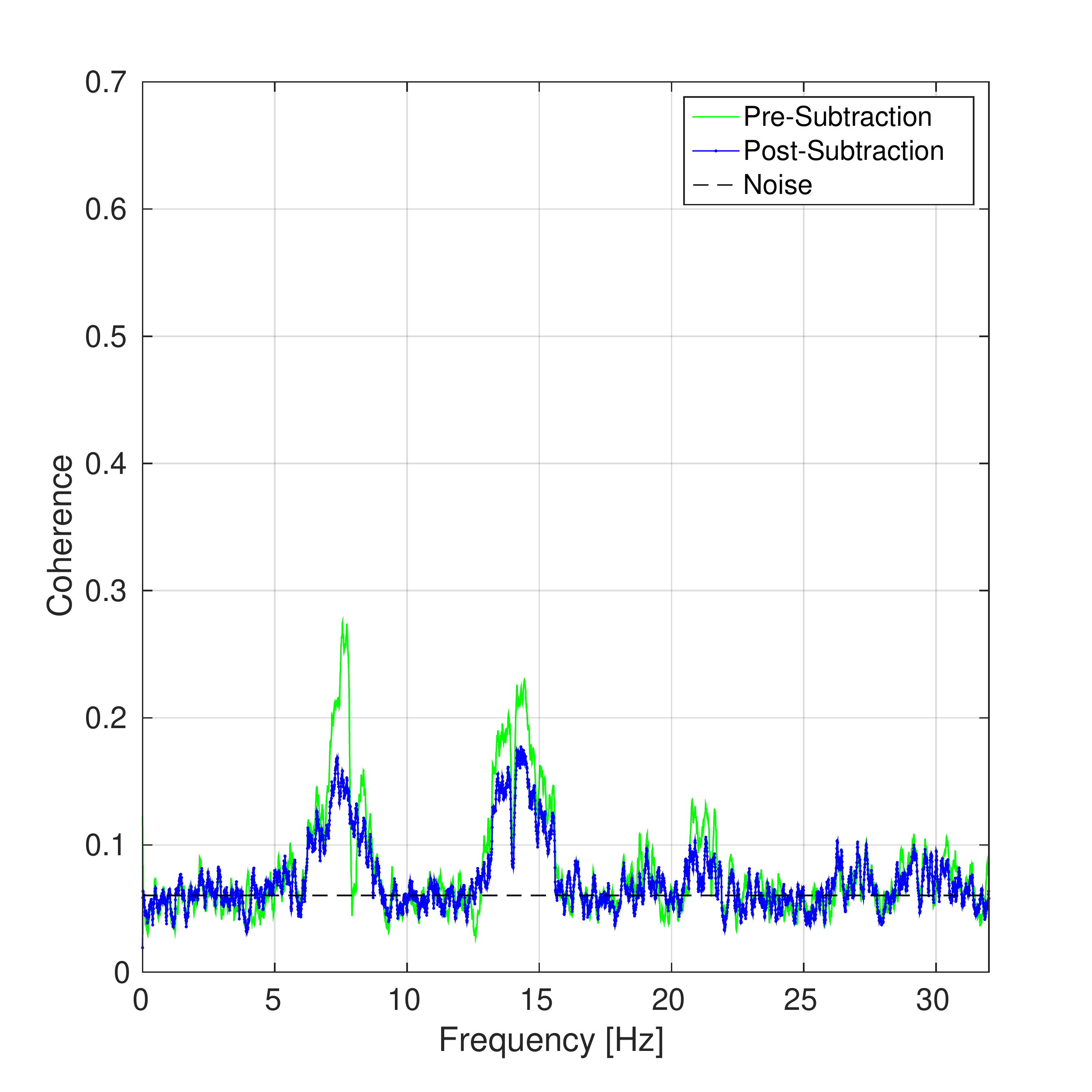} 
 KAGRA/Villa Cristina \hspace{0.3in} LIGO Livingston/Villa Cristina \hspace{0.3in} LIGO Hanford/Virgo
 \caption{On the left is the coherence between the KAGRA and the Villa Cristina magnetometer before and after Wiener filter subtraction.
For the KAGRA network, one KAGRA magnetometer is used as the target sensor while the second KAGRA magnetometer is the witness sensor.
For the Villa Cristina network, we use the Villa Cristina magnetometer as the target sensor and the orthogonal sensor, Poland Hylaty, and Colorado Hugo magnetometers as witness sensors.
 There is no measurable remaining peak from the Schumann resonances. 
 In the middle is the same between LIGO Livingston and the Villa Cristina magnetometers.
 For the LIGO Livingston network, we use the orthogonal sensor and Colorado Hugo, while for the Villa Cristina network, we use the orthogonal sensor and Poland Hylaty.
  Due to the distance from the witness sensors, some coherence remains.
 On the right is the same between LIGO Hanford and Virgo magnetometers.
 For the Virgo network, we use the orthogonal sensor, Poland Hylaty, and Patagonia, while for the LIGO Hanford network, we use the orthogonal sensor, LIGO Livingston, and Colorado Hugo.  
 }
 \label{fig:subtraction_coh}
\end{figure}

Wiener filtering relies on using the correlation of ``target'' and ``witness'' sensors, and witness sensors amongst one another, to remove noise common to the time series.
Sensors that have noise that is desired to be reduced are known as ``target'' sensors, which in this study are magnetometers but could be a gravitational-wave strain channel, for example.
The sensors that contain noise seen in the target sensor that are used for subtraction purposes are known as ``witness'' sensors.
We first use a Wiener filter to reduce a magnetometer's auto power spectral density. 
We then determine the level of reduction of correlated noise in two magnetometers, which is important for SGWB searches.
We made an attempt at global magnetometer subtraction in \cite{CoCh2016}, predominantly using previously deployed magnetometers to construct a toy-model gravitational-wave network. 
In this analysis, with the previously deployed magnetometers and dedicated on-site measurements, we provide results we can expect for on-site magnetometer based subtraction with a realistic network of magnetometers.

In our previous study \citep{CoCh2016}, we argued that it would be beneficial to measure the coherence between quiet magnetometers stationed near gravitational-wave detectors.
In this analysis, we use these magnetometers, in addition to dedicated measurements at the gravitational-wave detectors, to perform subtraction with realistic levels of local magnetic noise of correlated magnetic noise in gravitational-wave detectors.
Figure~\ref{fig:subtraction_auto} shows the ratio of the auto power spectral density before and after Wiener filter subtraction using a few different target and witness sensors.
For the first example, one KAGRA magnetometer is used as the target sensor while the second KAGRA magnetometer is the witness sensor.
In the second example, we use the Villa Cristina magnetometer as the target sensor and the sensor with the highest coherence in each frequency bin as the witness. Finally, we use the Villa Cristina magnetometer as the target sensor and all available sensors as witnesses.
The witness sensors in this test include the local orthogonal sensor, Poland Hylaty, and Colorado Hugo.
Using the entire network improves the subtraction by about 10\% above and beyond using the best channel only.
This result adds evidence to the notion that magnetic correlations are significant over the entire Schumann band. 

We now compute the correlation between magnetometers, both before and after Wiener filter subtraction, which is the metric most appropriate for searches for a SGWB.
We can use the available magnetometers as a proxy for a 2-detector gravitational-wave interferometer network.
In Figure~\ref{fig:subtraction_coh}, we show the coherence between the pairs of magnetometers we consider in this analysis.
On the left is the coherence between the KAGRA and Villa Cristina magnetometers before and after the subtraction to measure the effect that the Wiener filtering has had on the correlations.
For the KAGRA network, one KAGRA magnetometer is used as the target sensor while the second KAGRA magnetometer is the witness sensor.
For the Villa Cristina network, we use the Villa Cristina magnetometer as the target sensor and the orthogonal sensor, Poland Hylaty, and Colorado Hugo magnetometers as witness sensors.
There is no measurable remaining peak from the Schumann resonances. 
In the middle of Figure~\ref{fig:subtraction_coh} is the coherence between LIGO Livingston and Villa Cristina.
For the LIGO Livingston network, we use the orthogonal sensor and Colorado Hugo, while for the Villa Cristina network, we use the orthogonal sensor and Poland Hylaty as witness sensors.
We find that the coherence is reduced to the level of the expected noise floor in this case.
To test the permanent stations, we show the coherence between the LIGO Hanford and Virgo magnetometers before and after the subtraction on the right of Figure~\ref{fig:subtraction_coh}.
 For the Virgo network, we use the orthogonal sensor, Poland Hylaty, and Patagonia, while for the LIGO Hanford network, we use the orthogonal sensor, LIGO Livingston, and Colorado Hugo.  
 We achieve a reduction in coherence of about 60\%.
 In this case, some residual coherence visibly remains, mostly due to the distances between the target and witness sensors in the Virgo network.
 
\section{Conclusion}

In this paper, we have described magnetometer measurements at various gravitational-wave detector sites.
We computed optimal filters to perform subtraction between magnetometers.
We achieved subtraction near the level expected from an uncorrelated time series.
This shows that magnetometers near to the interferometers can effectively subtract magnetic noise with Wiener filtering. 
Going forward, it will be important to compute magnetometer correlations with gravitational-wave detector data in order to measure the effect from the Schumann resonances. From there, subtraction using magnetometers can be performed. Bayesian techniques that aim to separate magnetic contamination from gravitational-wave signals in cross-correlation search statistics are also being developed in parallel to those presented in this paper. It is important to approach the issue of magnetic contamination with many different methods as it promises to be a significant problem for cross-correlation-based SGWB searches in the future.

\acknowledgments
The authors would like to thank Dr. Brian O'Reilly for a careful reading of an earlier version of the manuscript.
The authors gratefully acknowledge the LIGO Observatories, European Gravitational Observatory (EGO), and KAGRA for technical support.
The magnetic field measurement in Kamioka was supported by the Joint
Usage/Research Center program of Earthquake Research Institute, the
University of Tokyo.
The authors also would like to thank Christina Daniel, Margarita Vidreo, and Julia Kruk, who helped bury the LHO site magnetometers, both during the site selection process and the final installation.
MC was supported by the David and Ellen Lee Postdoctoral Fellowship at the California Institute of Technology.
AC is supported by the INFN Doctoral Fellowship at the University of Genova.
NC work is supported by NSF grant PHY-1505373.
ET is supported through ARC FT150100281.
JK, AK and JM were supported by the National Science Center, Poland, under grant 2012/04/M/ST10/00565.

\bibliographystyle{unsrt}
\bibliography{references}

\begin{thebibliography}{10}

\bibitem{aligo}
{J Aasi et al.}
\newblock {Advanced LIGO}.
\newblock {\em Classical and Quantum Gravity}, 32(7):074001, 2015.

\bibitem{avirgo}
{F Acernese et al}.
\newblock {Advanced Virgo: a second-generation interferometric gravitational
  wave detector}.
\newblock {\em Classical and Quantum Gravity}, 32(2):024001, 2015.

\bibitem{AbEA2017h}
{The LIGO Scientific Collaboration}, {the Virgo Collaboration}, B.~P. {Abbott},
  R.~{Abbott}, T.~D. {Abbott}, F.~{Acernese}, K.~{Ackley}, C.~{Adams},
  T.~{Adams}, P.~{Addesso}, and et~al.
\newblock {GW170817: Implications for the Stochastic Gravitational-Wave
  Background from Compact Binary Coalescences}.
\newblock {\em ArXiv e-prints: 1710.05837}, October 2017.

\bibitem{AbEA2009}
BP~Abbott, R~Abbott, F~Acernese, R~Adhikari, P~Ajith, B~Allen, G~Allen,
  M~Alshourbagy, RS~Amin, SB~Anderson, et~al.
\newblock An upper limit on the stochastic gravitational-wave background of
  cosmological origin.
\newblock {\em Nature}, 460(7258):990--994, 2009.

\bibitem{AbEA2012s}
J.~Abadie, B.~P. Abbott, R.~Abbott, T.~D. Abbott, M.~Abernathy, T.~Accadia,
  F.~Acernese, C.~Adams, R.~Adhikari, C.~Affeldt, and {others}.
\newblock {Upper limits on a stochastic gravitational-wave background using
  LIGO and Virgo interferometers at 600--1000~Hz}.
\newblock {\em Phys. Rev. D}, 85:122001, Jun 2012.

\bibitem{AbEA2016b}
B.~P. et~al. Abbott.
\newblock {GW150914: Implications for the Stochastic Gravitational-Wave
  Background from Binary Black Holes}.
\newblock {\em Phys. Rev. Lett.}, 116:131102, Mar 2016.

\bibitem{TCS2013}
E~Thrane, N~Christensen, and RMS Schofield.
\newblock Correlated magnetic noise in global networks of gravitational-wave
  detectors: Observations and implications.
\newblock {\em Phys. Rev. D}, 87(12):123009, 2013.

\bibitem{TCS2014}
E.~Thrane, N.~Christensen, R.~M.~S. Schofield, and A.~Effler.
\newblock Correlated noise in networks of gravitational-wave detectors:
  Subtraction and mitigation.
\newblock {\em Phys. Rev. D}, 90:023013, Jul 2014.

\bibitem{LSC2007a}
{LIGO Scientific Collaboration}.
\newblock Upper limit map of a background of gravitational waves.
\newblock {\em Phys.~Rev.~D}, 76:082003, 2007.

\bibitem{LSC2007b}
{Abbott, B. et al.}
\newblock First cross-correlation analysis of interferometric and resonant-bar
  gravitational-wave data for stochastic backgrounds.
\newblock {\em Phys. Rev. D}, 76:022001, Jul 2007.

\bibitem{AaEA2014}
{Aasi, J. et al.}
\newblock {Improved Upper Limits on the Stochastic Gravitational-Wave
  Background from 2009-2010 LIGO and Virgo Data}.
\newblock {\em Phys. Rev. Lett.}, 113:231101, Dec 2014.

\bibitem{KoBi2017}
Izabela Kowalska-Leszczynska, Marie-Anne Bizouard, Tomasz Bulik, Nelson
  Christensen, Michael Coughlin, Mark Gołkowski, Jerzy Kubisz, Andrzej Kulak,
  Janusz Mlynarczyk, Florent Robinet, and Maximilian Rohde.
\newblock Globally coherent short duration magnetic field transients and their
  effect on ground based gravitational-wave detectors.
\newblock {\em Classical and Quantum Gravity}, 34(7):074002, 2017.

\bibitem{Schumann1}
W~Schumann.
\newblock {{\"U}ber die D{\"a}mpfung der elektromagnetischen Eigenschwingungen
  des Systems Erde — Luft — Ionosph{\"a}re}.
\newblock {\em Zeitschrift für Naturforschung A}, pages 250--252, 1952.

\bibitem{CoCh2016}
Michael~W Coughlin, Nelson~L Christensen, Rosario~De Rosa, Irene Fiori, Mark
  Gołkowski, Melissa Guidry, Jan Harms, Jerzy Kubisz, Andrzej Kulak, Janusz
  Mlynarczyk, Federico Paoletti, and Eric Thrane.
\newblock Subtraction of correlated noise in global networks of
  gravitational-wave interferometers.
\newblock {\em Classical and Quantum Gravity}, 33(22):224003, 2016.

\bibitem{GaNi2015}
Yu.P. Galuk, A.P. Nickolaenko, and M.~Hayakawa.
\newblock {Knee model: Comparison between heuristic and rigorous solutions for
  the Schumann resonance problem}.
\newblock {\em Journal of Atmospheric and Solar-Terrestrial Physics}, 135:85 --
  91, 2015.

\bibitem{Note1}
http://www.oa.uj.edu.pl/elf/index/projects3.htm.

\bibitem{KuKu2014}
A.~Kulak, J.~Kubisz, S.~Klucjasz, A.~Michalec, J.~Mlynarczyk, Z.~Nieckarz,
  M.~Ostrowski, and S.~Zieba.
\newblock Extremely low frequency electromagnetic field measurements at the
  {Hylaty} station and methodology of signal analysis.
\newblock {\em Radio Science}, 49(6):361--370, 2014.
\newblock 2014RS005400.

\bibitem{Note2}
http://www.lemisensors.com/?p=245.

\bibitem{Note3}
https://www.geo-metronix.de/mtxgeo/index.php/mfs-06e-overview.

\bibitem{Note4}
https://scientists.virgo-gw.eu/EnvMon/List/Trillium/Centaur.

\end{thebibliography}

\end{document}